\documentstyle[prd,aps,floats,epsfig,times]{revtex}
\draft
\newcommand{\be}{\begin{equation}}
\newcommand{\ee}{\end{equation}}
\newcommand{\deln}{$\Delta N_\nu\;$}
\newcommand{\epm}{$e^{\pm}\;$}
\def\ie{{\it i.e.},~}
\def\eg{{\it e.g.},~}
\def\4he{$^4$He}

\def\etal{{\it et al.}~}
\newcommand\la{\lower0.6ex\vbox{\hbox{$ \buildrel{\textstyle 
<}\over{\sim}\ $}}}
\newcommand\ga{\lower0.6ex\vbox{\hbox{$ \buildrel{\textstyle 
>}\over{\sim}\ $}}}

\begin{document}
\twocolumn[\hsize\textwidth\columnwidth\hsize\csname@twocolumnfalse\endcsname
\title{Precision Neutrino Counting}
\author{Gary Steigman}
\address{Departments of Physics and Astronomy, The Ohio State University,
Columbus, OH~~43210}
\date{{\today}}
\maketitle

\begin{abstract}

In the framework of the standard, hot big bang cosmological model 
the dynamics of the early evolution of the universe is controlled by 
the energy density of relativistic particles, among which neutrinos
play an important role.  In equilibrium, the energy density contributed
by one flavor of relativistic neutrinos is 7/8 of that of the cosmic 
background radiation (CBR) photons.  As the universe expands and cools, 
neutrinos decouple and their subsequent contribution to the energy 
density is modified by the relative heating of the CBR photons when 
electron-positron pairs annihilate.  The small corrections to the 
post-\epm annihilation energy density of the standard model neutrinos 
due to incomplete decoupling and finite-temperature QED effects are 
reviewed (correcting an error in the literature) and extended to 
account for possible additional relativistic degrees of freedom whose 
presence might modify the predictions of primordial nucleosynthesis 
and of the predicted CBR anisotropies. 

\end{abstract}

 \vskip 2pc]

\section{Introduction}

During the early evolution of the universe electroweak processes
establish and maintain equilibrium between the standard model
neutrinos ($e$, $\mu$, and $\tau$) and the cosmic background 
radiation (CBR) \cite{wein} - \cite{kt}.  In equilibrium the 
phase space distributions are Fermi-Dirac for the neutrinos and 
Bose-Einstein for the CBR photons and the temperatures are equal 
($T_{\gamma} = T_{\nu} \equiv T_{\nu_{e}} = T_{\nu_{\mu}} = 
T_{\nu_{\tau}}$).  The energy densities are related by
\be
\rho_{\nu} \equiv \rho_{\nu_{e}} = \rho_{\nu_{\mu}} = \rho_{\nu_{\tau}} 
= {7 \over 8}\rho_{\gamma}.
\ee
During early, radiation dominated (RD) epochs the universal expansion 
rate is uniquely determined by the energy density in relativistic 
particles.  Prior to \epm annihilation, at a temperature of a few MeV, 
the energy density receives its contributions from CBR photons, \epm 
pairs, and three flavors of neutrinos,
\be
\rho = \rho_{\gamma} + \rho_{e} + 3\rho_{\nu} = {43 \over 8}\rho_{\gamma}.
\label{rho0}
\ee
At this time ($T \sim$ few MeV) the neutrinos are beginning to decouple
from the photon -- \epm plasma and the neutron to proton ratio, which 
is key to the primordial abundance of \4he, is decreasing.  As a result, 
the predictions of primordial nucleosynthesis depend sensitively on the 
early expansion rate. During these RD epochs the age and the energy 
density are related by ${32\pi \over 3}G\rho t^{2} = 1$, so that the 
age of the universe is known once the temperature is specified,
\be
t~T_{\gamma}^{2} = 0.738~{\rm MeV^{2}~s}.
\label{tt}
\ee 

In many extensions of the standard models of cosmology and of particle 
physics there can be ``extra" energy density contained in new particles 
or fields, $\rho_{X}$.  When $X$ behaves like radiation, that is when its 
pressure and energy density are related by $p_{X} = {1 \over 3}\rho_{X}$, 
it is convenient to account for this extra energy density by normalizing 
it to that of an equivalent neutrino flavor 
\cite{ssg},
\be
\rho_{X} \equiv \Delta N_{\nu}\rho_{\nu} = 
{7 \over 8}\Delta N_{\nu}\rho_{\gamma}.
\label{deln}
\ee
In the following it is assumed that $X$ has decoupled prior to \epm
annihilation and its temperature (or, equivalently, its number density) 
is used to define a ``comoving volume".  For each such ``neutrino-like"
particle (\ie a two-component fermion), if $T_{X} = T_{\nu}$, then 
\deln = 1; if $X$ is a scalar, \deln = 4/7.  However, it may well be 
that $X$ has decoupled earlier in the evolution of the universe and 
has failed to profit from the heating when various particle-antiparticle 
pairs annihilated (or unstable particles decayed).  In this case, the 
contribution to \deln from each such particle will be $< 1$ ($< 4/7$).  
Since we are interested in the {\it deviations} of $T_{\gamma}$ and 
$T_{\nu}$ from $T_{X}$ when \epm pairs annihilate, for convenience 
and without loss of generality, we may define $T_{X} \equiv T_{\nu}$ 
prior to $e^{\pm}$ annihilation.  In the presence of this extra component, 
the pre-\epm annihilation energy density in eq.~\ref{rho0} is modified to,
\be
\rho = {43 \over 8}(1 + {7\Delta N_{\nu} \over 43})\rho_{\gamma}.
\label{rhox}
\ee
The extra energy density speeds up the expansion of the universe so that 
the right hand side of the time-temperature relation in eq.~\ref{tt} is 
smaller by the square root of the factor in parentheses in eq.~\ref{rhox}.

\section{Completely Decoupled Neutrinos}

For standard model neutrinos the electroweak interaction rates drop
below the universal expansion rate when the universe is less than
one second old and the temperature is a few MeV.  The ``standard", 
zeroth-order approximation that the neutrinos are completely decoupled 
prior to $e^{\pm}$ annihilation is sufficiently accurate for most 
cosmological applications.  In this approximation the CBR photons 
get the full benefit of the energy/entropy from the annihilating 
\epm pairs, maximizing their heating relative to the decoupled 
neutrinos (and to any other decoupled particles, such as $X$).  
After $e^{\pm}$ annihilation,
\be
T_{\nu} = T_{X} = ({4 \over 11})^{1/3}T_{\gamma} = 0.7138T_{\gamma}. 
\ee
In this approximation the Fermi-Dirac phase space distributions of the
decoupled neutrinos is preserved as the universe expands and $\rho_{\nu_{e}} 
= \rho_{\nu_{\mu}} = \rho_{\nu_{\tau}} \equiv \rho_{\nu}$ (and $\rho_{X} = 
\Delta N_{\nu}\rho_{\nu}$), where
\be
{\rho_{\nu} \over \rho_{\gamma}} = {7 \over 8}({T_{\nu} \over 
T_{\gamma}})^{4} = {7 \over 8}({4 \over 11})^{4/3} = 0.2271,
\ee
and
\be
{\rho_{X} \over \rho_{\gamma}} = {7 \over 8}({T_{X} \over T_{\gamma}})^{4}
\Delta N_{\nu} = {7 \over 8}({4 \over 11})^{4/3}\Delta N_{\nu}.
\ee
In this zeroth-order, full decoupling approximation, the post-\epm
annihilation relativistic ($R$) energy density is,
\be
{\rho_{R}^{0} \over \rho_{\gamma}} = 1 + {7 \over 8}({4 \over 11})^{4/3}N_{\nu} 
= 1.6813(1 + 0.1351\Delta N_{\nu}),
\ee
where $N_{\nu} \equiv 3 + \Delta N_{\nu}$.  As long as the universe 
remains radiation dominated, the age and the photon temperature are 
simply related by,
\be
t^{0}T_{\gamma}^{2} = 1.32(1 + 0.1351\Delta N_{\nu})^{-1/2}~{\rm MeV^{2}~s}.
\ee
Both $\rho_{R}$ and the time--temperature relation play important
roles in establishing the amplitudes and angular scales of the CBR
anisotropies and any corrections to these zeroth-order results may 
bias the interpretation of the precise data from current ground-based
and future space-based telescopes.

\section{Partially Coupled Neutrinos}

It has long been known that the neutrinos are not entirely decoupled 
from the electron-positron-photon plasma during \epm annihilation 
\cite{dicus}.  As a result, the neutrinos get to share some of the 
annihilation energy with the photons and the {\it relative} heating 
of the photons is reduced from the zeroth-order approximation estimate.  
Furthermore, since the weak interactions coupling the neutrinos and 
the electrons are energy dependent, the neutrino phase space distributions 
are distorted by preferential heating of the higher energy/momentum 
neutrinos and the resulting distributions are no longer Fermi-Dirac.  
In a series of increasingly detailed calculations, many authors have 
tracked this evolution \cite{dicus} - \cite{gg}.  Although there are 
small differences in the quantitative results which may be traced to 
the differing approximations, the overall agreement among them is 
excellent.  The results of the detailed and extensive calculations 
of Gnedin \& Gnedin~\cite{gg}, which cover some seven orders of 
magnitude in the neutrino momentum, are adopted here.

Recall from eq.~7 that in the fully decoupled approximation, 
$\rho_{\nu}/\rho_{\gamma} = 0.2271$.  Since the electron neutrinos
participate in charged-current as well as in neutral-current weak
interactions, they remain coupled longer than do the $\mu$ or $\tau$ 
neutrinos and, thereby, are heated more.  In following the decoupling 
carefully, Gnedin \& Gnedin find,
\be
{\rho_{\nu_{e}} \over \rho_{\gamma}} = 0.2293 = 
{7 \over 8}({4 \over 11})^{4/3}\times 1.0097,
\ee
while
\be
{\rho_{\nu_{\mu}} \over \rho_{\gamma}} = 
{\rho_{\nu_{\tau}} \over \rho_{\gamma}} = 0.2285 = 
{7 \over 8}({4 \over 11})^{4/3}\times 1.0061.
\ee
Further, because the energy/entropy from the \epm is now being
shared by the neutrinos, the heating of the photons relative to
the $X$ (assumed to be fully decoupled prior to annihilation) is
reduced, so that $T_{X} = 0.7144T_{\gamma}$ (compared to the 
zeroth-order estimate in eq.~6) and,
\be
{\rho_{X} \over \rho_{\gamma}} = 
{7 \over 8}({T_{X} \over T_{\gamma}})^{4}\Delta N_{\nu} = 
{7 \over 8}({4 \over 11})^{4/3}\Delta N_{\nu}\times 1.0036.
\ee

We are now in a position to combine the post-\epm annihilation
energy densities of the photons along with that of the neutrinos 
(and possible $X$s) to find the relativistic energy density in 
the incompletely decoupled ($ID$) neutrino approximation.  This 
result may be written in analogy with eq.~9 by replacing $N_{\nu}$
with $N_{\nu}^{ID}$,
\be
{\rho_{R}^{ID} \over \rho_{\gamma}} \equiv 1 + 
{7 \over 8}({4 \over 11})^{4/3}N_{\nu}^{ID},
\ee
where
\be
N_{\nu}^{ID} = 3.022 + 1.0036\Delta N_{\nu},
\ee
or
\be
N_{\nu}^{ID} = N_{\nu} + (0.022 + 0.0036\Delta N_{\nu}).
\label{nnuid}
\ee
Thus, for the standard model case of $N_{\nu} = 3$, the 
post-\epm annihilation energy density corresponds to 3.02 
``equivalent" neutrinos.  Although all authors (\cite{dicus} 
- \cite{gg}) agree quantitatively with this \deln = 0 correction, 
a {\it different} value (3.03) appears in the Lopez \etal paper 
\cite{ldht}.  It is difficult to identify the source of the 
Lopez \etal correction, $\delta N_{\nu}^{ID} = 0.03$, which 
differs from that found earlier by some of the same authors 
\cite{dt,fdt}.  The Lopez \etal value, modified by the QED 
effect to be addressed next ($\delta N_{\nu}^{QED} = 0.01$), 
seems to have propagated in the recent literature (see, \eg 
\cite{hann,bean}) and, indeed, 3.04 is the recommended default 
value for $N_{\nu}$ in the CMBFAST code of Seljak \& Zaldarriaga 
\cite{CMBFAST}, widely utilized in analyses of the CBR fluctuation 
spectra.  The second term on the right hand side of eq.~\ref{nnuid}, 
the difference between $N_{\nu}^{ID}$ and $N_{\nu}$, is the 
incomplete decoupling correction to the standard model result, 
{\it including} the effect of extra relativistic energy.  
However, there is another effect which, while small, is 
the same order of magnitude as this correction for incomplete 
decoupling.

\section{QED Correction}

In the zeroth-order approximation, the $\gamma$ -- \epm plasma is 
treated as a gas of free, non-interacting particles prior to \epm 
annihilation.  When finite temperature QED corrections are included 
(\cite{heck,lt}), the energy density and pressure of the $\gamma$ 
-- \epm plasma are reduced.  As a result, when \epm pairs annihilate 
they actually have less entropy to share with the photons (and the 
incompletely decoupled neutrinos), than would be estimated neglecting 
this QED correction.  Accounting for this reduction leads to fewer 
CBR photons (and neutrinos) in the post-\epm annihilation comoving 
volume, corresponding to a lower photon temperature.  To quantify 
this effect (\cite{heck,lt}) it is convenient to compare the 
post-\epm annihilation photon temperature to that of the fully 
decoupled $X$ particles.
\be
(T_{X}/T_{\gamma})_{QED} = (4/11)^{1/3}(1 + 9.6\times 10^{-4}).
\ee
Thus, in comparing the total energy density in relativistic
particles to that in CBR photons alone, the {\it relative}
contribution of the neutrinos (and $X$s) is {\it enhanced}
with respect to the contributions in the zeroth-order 
approximation (eq.~9) and in the incompletely decoupled
approximation (eq.~14).  To account for the finite temperature
QED correction, the $(4/11)^{4/3}$ factors on the right 
hand sides of eqs.~9 and 14 should be replaced by
$(T_{X}/T_{\gamma})^{4}_{QED}$, leading to a new, 
``effective" number of equivalent neutrinos $N_{\nu}^{eff}$,
\be
N_{\nu}^{eff} \equiv 
(11/4)^{4/3}(T_{X}/T_{\gamma})^{4}_{QED}N_{\nu}^{ID} 
= 1.0038N_{\nu}^{ID},
\ee
so that 
\be
N_{\nu}^{eff} = 3.034 + 1.0074\Delta N_{\nu},
\ee 
or
\be
N_{\nu}^{eff} - N_{\nu} = 0.034 + 0.0074\Delta N_{\nu}.
\label{neff}
\ee
This {\it total correction} to the standard model, 
zeroth-order approximation, $N_{\nu} = 3 + \Delta N_{\nu}$, 
is shown in Figure 1 as a function of any extra energy 
density, measured by $\Delta N_{\nu}$.

\section{Discussion}

Prior to \epm annihilation the total energy density in the
standard model ($N_{\nu} = 3$) is given in terms of the CBR 
energy density by eq.~2.  In the presence of extra (relativistic) 
energy density, \deln $> 0$, the $\rho_{R}-\rho_{\gamma}$ 
relation is modified to that in eq.~5.  After \epm annihilation 
the relativistic energy density is shared by photons and decoupled 
neutrinos (and $X$s) and the total energy density is modified 
by the relative heating of the photons, neutrinos, and $X$s.  
When careful account is taken of the incomplete decoupling 
of the neutrinos during \epm annihilation, as well as of the 
finite temperature QED effects on the equation of state of 
the pre-annihilation plasma, the total energy density
($\rho_{R} \equiv \rho_{R}^{eff}$), normalized to the CBR 
photon energy density, is
\be
{\rho_{R} \over \rho_{\gamma}} = 1 + 
{7 \over 8}({4 \over 11})^{4/3}N_{\nu}^{eff},
\ee
where eq.~\ref{neff} provides the connection between 
$N_{\nu}^{eff}$ and $N_{\nu}$.  For the standard model case
of $N_{\nu} = 3$, the correction to the post-\epm annihilation
energy density, 0.034, is small; but, for \deln $> 0$, 
this correction term grows.  The new term, proportional 
to $\Delta N_{\nu}$, will dominate the correction for \deln 
$\ga 4.5$, and should not be ignored if precision requires 
that the first term be included.  The difference between the 
zeroth-order energy density (eq.~9) and the corrected one 
(eq.~21) is also shown in Figure 1.  It is interesting
to note that in a joint BBN + CBR analysis of the constraints
on $N_{\nu}^{eff}$, Hansen \etal (\cite{hmmmp}) recommend a
standard model (\deln = 0) value of $N_{\nu}^{eff} = 3.034$, 
in excellent agreement with that presented here, but they 
make no mention of the additional corrrection when \deln 
differs from zero.

\begin{figure}[htbp]
\begin{center}
\epsfig{file=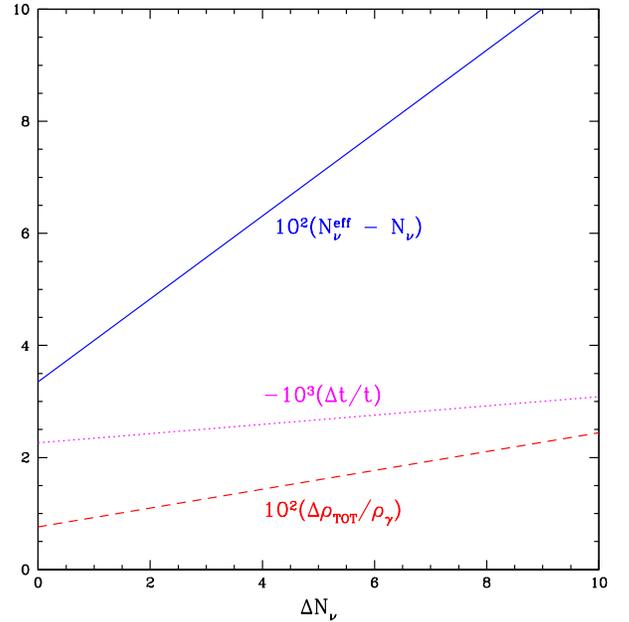,width=3.5in}
	\caption{The solid curve is the correction to
	the zeroth-order result for the equivalent number of
	neutrinos as a function of the ``extra" equivalent
	number of neutrinos.  The dotted curve shows the 
	fractional change in the post-\epm annihilation age
	when the universe is radiation dominated.  The dashed
	curve is the difference between the corrected, and 
	zeroth-order radiation energy densities in units of 
	the CBR photon energy density.}
 	\label{fig1}
\end{center}
\end{figure}
Since the post-\epm annihilation expansion rate is determined by 
$\rho_{R}$ (as long as the universe remains RD), the age -- temperature
relation will be modified from its zeroth-order form (eq.~10), 
\be
{t^{0} \over t} = ({1 + 0.1351(N_{\nu}^{eff} - 3) \over
1 + 0.1351(N_{\nu} - 3)})^{1/2}.
\ee
The fractional change in the age of the universe ($\Delta t /t \equiv 
(t - t^{0})/t$) is shown in Figure 1.  Since the CBR fluctuation
spectrum is being painted on the microwave sky during those epochs
when the universe is making the transition from radiation to matter
dominated, modifications to the zeroth-order expressions for the
expansion rate and the radiation energy density will effect the 
details of the resulting anisotropies.  

As cosmology enters this new era of precision science, it may be 
necessary to account for even the very small changes from the 
zeroth-order approximation summarized here.  For example, Lopez 
\etal (\cite{ldht}) suggest that measurements accurate to $\delta 
N_{\nu} \approx 0.03$, a $\sim 1$\% determination, will be possible.  
At present, such precision seems a distant dream.  For example, 
for \deln = 0, the effect on the BBN-predicted helium abundance 
is very small (\cite{fdt,lt}), $\delta$Y$_{\rm P}$ = 1.5$\times 
10^{-4}$, a correction buried in the overall uncertainty in the 
BBN prediction ($\ga 4\times 10^{-4}$~\cite{bnt}).  The small 
speed up in the post-\epm expansion rate will modify the BBN-predicted 
abundances of the other light elements, but here, too, the changes 
(for \deln = 0) are overwhelmed by the current theoretical 
uncertainties (see, \eg \cite{bnt}).  As for the CBR angular 
fluctuations, there are degeneracies between $N_{\nu}^{eff}$ 
and many other cosmological parameters, leading at present 
to only very weak constraints (\cite{kssw,hann2,hmmmp}), 
$N_{\nu}^{eff} \la 7 - 17$.  At this stage, differences between 
$N_{\nu}^{eff}$ and $N_{\nu}$ at the 1\% level are beyond 
our grasp.  We can only hope that our experimental colleagues 
will rise to the challenge and provide data of such exquisite 
precision that the corrections reviewed and summarized here 
will be important.

\acknowledgments
For valuable discussions or email exchanges I wish to thank Steen
Hannestad, Andrew Heckler, Uros Seljak and Matias Zaldarriaga.  I
also acknowledge the support of the DOE through grant DE-FG02-91ER40690.


\end{document}